\documentclass{article}
\usepackage[utf8]{inputenc}
\pdfoutput=1

\title{A Review of Quantum and Hybrid Quantum / Classical Blockchain Protocols}
\author{M. Edwards$^1$, A. Mashatan$^2$, and S. Ghose$^{3,1}$\\
$^1$Institute for Quantum Computing, University of Waterloo, Canada\\
$^2$ School of Information Technology Management, Ryerson University, Toronto, Canada\\
$^3$Department of Physics and Computer Science, Wilfrid Laurier University, Waterloo, Canada}
\date{}

\usepackage{natbib}
\usepackage{graphicx}
\usepackage{amsmath}
\usepackage{amsfonts}
\usepackage{amssymb}
\usepackage{mathtools}
\usepackage{rotating}
\usepackage{longtable}
\usepackage{multicol}
\usepackage{fontawesome}

\usepackage{listings}
\usepackage{color}

\definecolor{dkgreen}{rgb}{0,0.6,0}
\definecolor{gray}{rgb}{0.5,0.5,0.5}
\definecolor{mauve}{rgb}{0.58,0,0.82}

\begin{document}

\begin{center}

{\LARGE A Review of Quantum and Hybrid Quantum / Classical Blockchain Protocols}

\vspace{5mm}

M. Edwards$^1$, A. Mashatan$^2$, and S. Ghose$^{3,1}$\\
$^1$Institute for Quantum Computing, University of Waterloo, Canada\\
$^2$ School of Information Technology Management, Ryerson University, Toronto, Canada\\
$^3$Department of Physics and Computer Science, Wilfrid Laurier University, Waterloo, Canada

\vspace{5mm}

To whom correspondence should be addressed:\\
msedwards@uwaterloo.ca \\
519-998-5843
\end{center}

\pagebreak

\section{Abstract}

Blockchain technology is facing critical issues of scalability, efficiency and sustainability. These problems are necessary to solve if blockchain is to become a technology that can be used responsibly. Useful quantum computers could potentially be developed by the time that blockchain will be widely implemented for mission-critical work at financial and other institutions. Quantum computing will not only cause challenges for blockchain, but can also be harnessed to better implement parts of blockchain technologies including cryptocurrencies. We review the work that has been done in the area of quantum blockchain and hybrid quantum-classical blockchain technology and discuss open questions that remain.

\section{Introduction}

Quantum blockchain technology is one of the areas of research in the rapidly growing field of quantum cryptography [1]. Quantum cryptographic schemes make use of quantum mechanics in their designs. This enables such schemes to rely on presumably unbreakable laws of physics for their security. Many quantum cryptography schemes are information-theoretically secure, meaning that their security is not based on any non-fundamental assumptions. In the design of blockchain systems, information-theoretic security is not proven. Rather, classical blockchain technology typically relies on security arguments that make assumptions about the limitations of attackers' resources. 

Blockchain and distributed ledger technologies have applications in many industries, most notably in the financial industry. The financial applications of blockchain technologies include cryptocurrencies, insurance and securities issuance, trading and selling. Non-financial applications of blockchain technology have been identified for the music industry, decentralized IoT, anti-counterfeit solutions, internet applications and decentralized storage, to name a few. In recent years, blockchain projects have attracted massive attention in these industries [2].

Despite being a relatively new technology, blockchain has made significant waves in a number of important industries in a very short time. The two most known instances of blockchain technologies are Bitcoin [3] and Ethereum [4], which are the core of modern cryptocurrencies. Ethereum's focus on smart contracts has made it a valuable tool for decentralizing numerous industries.

The philosophical implications of decentralized consensus technologies are far-reaching. Atzori suggested in 2015 that all of society might be restructured by the blockchain, and that "the decentralization of government services through permissioned blockchains is possible and desirable" [5].

In this paper we review work that introduces quantum cryptographic methods to blockchain technology. We discuss the potential impact and risk associated with blockchain technology and how the proposed quantum cryptographic methods attempt to address these risks.

\section{Blockchain Background}

Before delving into quantum and hybrid quantum-classical blockchain crytography schemes, we will provide a brief summary of the core mechanisms in blockchain technology. The National Institute for Standards and Technology (NIST) describes blockchain technology in the following way:

\begin{quote}
    Blockchains are tamper evident and tamper resistant digital ledgers implemented in a distributed fashion (i.e., without a central repository) and usually without a central authority (i.e., a bank, company, or government). [6]
\end{quote}

We will focus primarily on Ethereum's blockchain implementation in the following sections since Ethereum's smart contracts have inspired interesting work in theoretical quantum blockchain design. Ethereum was introduced by Vitalik Buterin in 2013 [4]. The most important feature of Ethereum is arguably its Turing-complete scripting language for smart contracts. Ethereum shares many basics with other blockchain implementations. Here, we will summarize the elements of Ethereum  that are most relevant to the work that we review [4]. We begin with the basics that are shared by Ethereum and Bitcoin.

\subsection{The Ledger}

The distributed ledger of a blockchain cryptocurrency maintains the ownership and status of all existing coins. The ledger is made up of a chain of blocks. The chain is composed of the blocks' references to one another. Any valid block's header contains a hash of the header of the previous block in the chain. Each block typically also contains a timestamp, nonce, and list of transactions.

\vspace{5mm}
\begin{center}
    \includegraphics[scale=0.27]{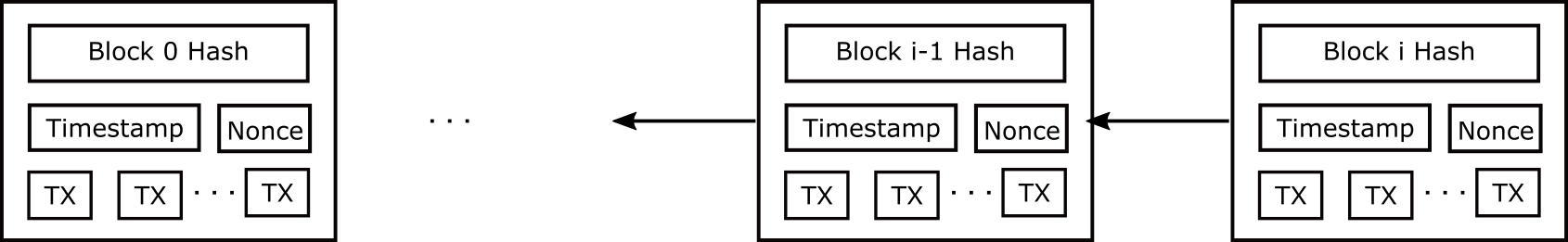}
    FIG. 1. Chain of Blocks
\end{center}
\vspace{5mm}

When a transaction occurs, the current ledger state is mutated by a function that takes the original state $S_0$ and the transaction $TX$, and outputs the next state $S_1$ or an error $E$. Here, and throughout this paper, $\leftarrow$ represents a transition of a state. Note that in this case, the state is a purely classical data structure. However, in later sections the same notation will be used to denote transitions between quantum states.

\vspace{5mm}
\[
S_1 or E \leftarrow Apply(S_0, TX)
\]
\vspace{5mm}

In Bitcoin, a ledger's state is composed of all Unspent Transaction Outputs ($UTXO$), or simply all of the coins that have been mined but not spent. Each coin has a 20-byte cryptographic public key which contains information about its owner and its denomination.

A transaction requires references to each $UTXO$ involved and the cryptographic signatures produced by the $UTXO$ owners' private keys.

\subsection{Proof of Work}

To achieve the decentralization of the ledger, a consensus system must be introduced. The goal of the consensus system is to ensure that everyone agrees on the validity of the transactions that have led to the ledger's state and their order. There are several consensus systems that are in use today, including proof-of-work, proof-of-stake, proof-of-burn and more. The most ubiquitous is proof-of-work.

Bitcoin's proof-of-work based system requires that users attempt to publish transactions constantly. These transactions are published in packaged groups of a fixed size (1 MB in the case of Bitcoin) called blocks. In addition to a list of transactions, a block contains a timestamp, one time use block id or nonce, and a hash of the header of the last most-recent block that contributed to the ledger. Hence each block maintains a reference to the block that came before it, and the blocks form a chain as they are published which reflects the order of their publications in time.

In order for a block to be accepted, its proof of work must be valid. The validity condition for Bitcoin block is that its double-SHA256 hash is less than a dynamically adjusted cutoff when interpreted as an integer. A SHA256 hash is a completely unpredictable result of a pseudorandom function. So, in order to create a valid block the hash function must be run an arbitrary number of times until a valid output randomly occurs. Therein lies the work that must be done to generate a proof-of-work, and the incredible overhead in computational resources that is encouraged by proof-of-work blockchains.

The time required to generate a valid hash is fundamental to the consensus system that is employed. If an attacker attempts to move money in a way that conflicts with the ownership of coins as a result of a transaction record already accepted by the ledger, the attack is simply rejected. However, an attacker can try to fabricate a block which points to a valid block that was published \textit{before} the block containing the transaction which changed the ownership of the desired coins. In this case, the attacker will be required to generate a new valid proof-of-work. While the attacker is occupied doing this, it is assumed that many other miners are continuing to publish blocks that point to the latest legitimate block. The rule that is applied to weed out these attacks is simply that the longest valid chain is taken to be the truth. An attacker would therefore need to have more computing power than the rest of the network combined in order to outpace the speed of the network's publications and make his/her chain the longest. This is called a 51\% attack and would theoretically be successful.

\subsection{Proof of Stake}

Proof-of-stake schemes were introduced to address some of the issues with proof-of-work [7]. Ethereum is currently in the process of switching to a proof-of-stake scheme. The mining power available to a miner in a proof-of-stake scheme is proportional to the number of coins owned by the miner. Hence, they are limited to mining a number of blocks that is proportional to their stake in the cryptocurrency ecosystem. This offloads the miners' consumption of electrical energy resources to currency resources that are more internal to the blockchain.

A driving force behind the creation of proof-of-stake was the dynamic created by miners selling their coins to pay off their electrical bills. This movement of cryptocurrency out of the ecosystem has led to drops in cryptocurrency value.

Proof-of-stake schemes have less inherent risk than proof-of-work schemes. This is clearly illustrated by the proof-of-stake version of the 51\% attack. In a proof-of-stake blockchain, an attacker would need to have 51\% of the cryptocurrency in the ecosystem to make a successful 51\% attack. This would make it unappealing to attack the ecosystem, since destroying the security and validity of the system would risk invalidating the attacker's virtual fortune. This is a natural deterrent that does not exist in proof-of-work schemes, where any attacker with 51\% of the network's computing power can make a successful 51\% attack regardless of their stake in the ecosystem.

There are cons to any consensus algorithm. In the case of proof-of-stake, one problem is the explicit association of wealth with the power to influence events. While the scheme improves on proof-of-work in some ways, it still incentivizes competition and, similarly to evolutionary systems, rewards the "fittest" competitor. In this case, fitness is quantified by units of currency rather than computational capabilities.

\subsection{Smart Contracts}

One of Ethereum's most significant contributions to blockchain technology is the concept of autonomous smart contracts. The addition of smart contracts differentiates so-called "Blockchain 2.0" technology like Ethereum from "Blockchain 1.0" technology like Bitcoin [8]. Blockchain 2.0 technologies enable programmers to use autonomous agents, the smart contracts, as elements of distributed software applications called Distributed Applications (DApps).

The top level data structures in Ethereum's ledger are accounts, rather than coins. The ledger maintains each account's 20-byte public-key address, nonce, balance, contract code and storage.

There are two types of accounts. The first is Externally Owned Accounts, which are controlled by private keys. The second is Contract Accounts which are controlled by their contract codes. Externally owned accounts are similar to those used by Bitcoin, and can be used in transactions as described previously. Contract accounts are much more interesting. Contract accounts act as autonomouus agents which execute their contract code when sent messages. A contract account can be programmed to automatically read and write to its storage, send additional messages to other contract accounts, or create transactions. To cause a contract account to execute its code exacts a monetary price on the sender of the original message. This price is known as "gas" and is proportional to the complexity of the contract code. The money (Ether) provided in the original message is used as gas to "fuel" all contract code executions that result from the first contract's activation.

Contract code is written in a low level stack machine based bytecode language called Ethereum Virtual Machine (EVM) code. The language makes use of a stack, linear memory array and long term storage. The language is composed of a small instruction set that includes blockchain applicaiton specific instructions like $CALL$ which sends a message to a contract and $CREATE$ which creates a new contract.

The blocks used by Ethereum are very similar to those used by Bitcoin. Ethereum blocks contain all the information that a Bitcoin block does, with the addition of a copy of the most recent ledger state, the block number and a record of the mining difficulty for that block.
Ethereum does not fundamentally deviate from the typical proof-of-work consensus scheme.

\section{Quantum Coins}

A straightforward way to introduce quantum technology to the blockchain at the cryptocurrency level is to simply reference the many schemes for quantum money that have been defined since 1960 [9]. Bitcoin and Ether were described in section 3 as the representations of monetary value that are traded between parties through transactions. These coins have monetary value and cryptographically protected ownership records. Coins are one of the primitive data structures required to formulate a cryptocurrency blockchain.

\subsection{Public-Key Quantum Money}

In the case of public-key quantum money, the scheme takes advantage of superposition to ensure that when a quantum state is used as a coin no bad actor can duplicate the coin. An attacker cannot know which basis to measure each qubit of the quantum state in without knowing the secret key which was originally used to create the state. The attacker cannot learn the state without performing the correct measurement due to the no-cloning restriction.

The procedure to generate public-key quantum money is very straightforward, and was originally introduced in 1960 by Stephen Wiesner [9]. This paper arguably kicked off the field of quantum cryptography and directly inspired the design of BB84 quantum key distribution [10].

An algorithm for public-key quantum money generation is simply the following:

\begin{itemize}
    \item[1.] Generate two random bit strings $M$ and $N$ of length $l$
    \item[2.] Prepare a quantum state $|\$>=|0>^{\otimes l}$
    \item[3.] For each bit $i < l$:
        \begin{itemize}
            \item If $M_i = 0$ and $N_i = 0$, do nothing to the $i^{th}$ qubit
            \item If $M_i = 0$ and $N_i = 1$, rotate the $i^{th}$ qubit state to $|1>$
            \item If $M_i = 1$ and $N_i = 0$, rotate the $i^{th}$ qubit state to $|+>$
            \item If $M_i = 1$ and $N_i = 1$, rotate the $i^{th}$ qubit state to $|->$
        \end{itemize}
\end{itemize}

$M_i$ and $N_i$ are kept secret by the mint, and $|\$>$ is published as the quantum public key. In this case, only the mint has enough knowledge to verify the public key and no one can duplicate it.

\subsection{Binding Commitments}

Some of the mechanisms that were taken for granted in the description of classical blockchain technology are non-trivial to implement using quantum algorithms. For example, we described in section 3.1 that a transaction requires references to each $UTXO$ involved and the cryptographic signatures produced by the $UTXO$ owners' private keys. This information is necessary to validate the ownership of the coins involved in the transaction. Using his/her private key, the owner of the coins creates a digital signature so that other parties can verify that the transaction was indeed authorized by the owner of the private key and was not modified since. Using the corresponding public key and the signature, any party can verify the validity of the transaction without learning the private key. This is the basic premise of public-key cryptography.

A blockchain transaction using quantum money will still require references to each $UTXO$ involved and the cryptographic signatures produced by the $UTXO$ owners’ private keys in order to verify ownership. It makes sense to also require that a user who has committed coins to a transaction in good faith \textit{must} produce his/her signature when it is time for the transaction to be approved. This would make the creation of a transaction using quantum public-key money as coins a type of binding commitment.

Computationally binding commitment schemes between two parties are composed of two phases. The Commitment Phase allows one party to send the other party some information $c$ related to a message $m$ which does not give the receiver any information about $m$ itself. However, the act of sending $c$ binds the sender to provide the message $m$ in the second stage, the Open Phase. In the Open Phase, the sender transmits $m$ to the receiver and proves to the receiver that $m$ does indeed correspond to $c$ by providing a signature that "opens $c$ to $m$".

A classical definition of a computationally binding is the following from Unruh [11].

\vspace{5mm}

\textbf{Definition 1 (Classical-style binding)}
\textit{No algorithm A can output a commitment c and two signatures s, s' that open c to two different messages m and m'.}

\vspace{5mm}

Computationally binding commitment schemes have been studied and defined in the quantum setting [11, 20, 21, 22]. Interestingly, when the algorithm $A$ is allowed to be a quantum polynomial time algorithm, this definition was shown to be inadequate. While definition 1 holds for a particular classical-style binding commitment, Ambainis, Rosmanis, and Unruh showed that for this particular binding a quantum polynomial time algorithm $A$ employed by an adversary could open $c$ to any message that the adversary wished [12]. Therefore Unruh was motivated to define a different type of binding that was useful in the quantum case. The new binding property is demonstrated by a pair of quantum games.

Let $A$, $B$ be algorithms and $S$, $M$, $U$ be quantum registers. $V_c$ is a measurement which verifies that that $U$ opens $M$. $M_{ok}$ measures 
m in the computational basis if $ok = 1$.

The first game $Game_1$ consists of four steps:

\vspace{5mm}

\[
(S, M, U, c) \leftarrow A(1^{\gamma})
\]
\[
ok \leftarrow V_c(M, U)
\]
\[
m \leftarrow M_{ok}(M)
\]
\[
b \leftarrow B(1^{\gamma}, S, M, U)
\]

\vspace{5mm}

The second game $Game_2$ omits the measurement in step three but is otherwise the same:

\vspace{5mm}

\[
(S, M, U, c) \leftarrow A(1^{\gamma})
\]
\[
ok \leftarrow V_c(M, U)
\]
\[
b \leftarrow B(1^{\gamma}, S, M, U)
\]

\vspace{5mm}

A commitment scheme is "collapse-binding" iff for any quantum polynomial time valid adversary, $cAdv = |Pr[b=1: Game_1] - Pr[b=1: Game_2]|$ is negligible.

This essentially expresses that if an adversary $(A, B)$ provides a classical commitment $c$, there must be only one message they can open $c$ to. $A$ outputs a superposition of messages $M$ and a superposition of corresponding opening signatures $U$. $S$ is the adversary's state. The assertion that $|Pr[b=1: Game_1] - Pr[b=1: Game_2]|$ is negligible limits the value of $M$ to computational basis vectors for collapse-binding commitments. No quantum polynomial time algorithm $B$ should be able to distinguish between the value of $M$ whether $M$ is measured in the computational basis or not.

\subsection{Collapsing Hash Functions}

The games used to define the collapse-binding property of commitment schemes can also be applied to classify hash functions that are collapsing [11]. Assume $H$ is a one-to-one hash function.

\vspace{5mm}

\textbf{Definition 2 (Collapsing hash function - informal)} 
\textit{H is a collapsing hash function iff no quantum polynomial time algorithm B can distinguish between $Game_1$ and $Game_2$. An adversary is valid if A outputs a classical value c and a register M where H(m) = c.}

\vspace{5mm}

This game based definition was clarified and made mathematical by Fehr in 2018 [13].

\vspace{5mm}

\textbf{Definition 3 (Collapsing hash function - formal)}
\textit{A function H $\mathbb{X} \rightarrow \mathbb{Y}$ is $\in$(q)-collapsing if}

\[
cAdv[H](q) :=
\]
\[
\underset{S M C U}{sup} \delta_q(M, \overline{M} | \overline{C} U) \leq \in(q)
\]

\vspace{2mm}

\textit{for all q. The supremum is over all states $S M C U$ = $S$  $H(M)$  $C U$ with complexity $\leq q$.}

\vspace{5mm}

The collapsing property of a hash function is a counterpart of collision resistance. Unruh shows that the random quantum oracle is a collapsing hash [11] and so some hash function based commitment schemes are collapsing in the random oracle model. Unruh also showed that Merkle-Damgard hash functions are collapsing if their underlying compression algorithms are, which implies that SHA-2 is collapsing [14]. Czajkowski, et al. showed the same for Sponge hashes with certain conditions [15]. Sponge hash construction underlies SHA-3.

\subsection{Collision Free Quantum Money}

Collision free quantum money is a concept that was introduced by Lutomirski, et al. [16]. The premise is that a mint can not efficiently produce two coins with the same verification circuit, and so each coin made is unique. This is a step towards remedying the problem with Wiesner's public-key quantum money. In Wiesner's scheme, only the mint can verify the quantum public keys of minted coins. This is an important issue specifically in the context of blockchain, since the intention is specifically \textit{not} to have a centralized signing authority in a distributed system.

Let $L$ be a classical function that assigns a unique label to each exponentially small subset of a superset of elements. $L$ should also be as obscure and unstructured as possible. The procedure for generating collision free quantum money is the following.

\begin{itemize}
    \item Begin with an equal superposition over all n-bit strings.
    \item Compute $L$ into an ancilla register and measure that register to obtain a value $l$.
\end{itemize}

This procedure would have to be repeated exponentially many times to produce the same value $l$ twice. The quantum state will then be $|\$_l>$, an equal superposition of exponentially many terms with no clear relationship to one another.

\[
|\$_l> = \frac{1}{\sqrt{N_l}} \sum_{x s.t. L(x)=l} |x>
\]

Verification can be done using rapidly mixing Markov chains. Verification requires knowledge of a Markov matrix $M$ that will rapidly mix from any distribution over bit strings with the same $l$ to the uniform distribution of those same strings. No string with a different $l$ can be present in that final uniform distribution. Each update consists of a uniform random choice over $N$ update rules $P_i$. Each update rule is deterministic and invertible. Then any valid quantum money state will be a +1 eigenstate of $M$.

\[
M^r \dot{=} \sum_l |\$_l><\$_l|
\]

\[
M = \frac{1}{N} \sum^N_{i=1} P_i
\]

The verification procedure makes use of a unitary $U$.

\[
U = \sum_i P_i \otimes |i><i|
\]

The verification procedure itself is the following:

\begin{itemize}
    \item Introduce an ancilla in uniform superposition over all i.
    \item apply $U$.
    \item Measure the projector of the ancilla onto the uniform superposition.
    \item discard the ancilla.
\end{itemize}

The outcome 1 has a corresponding Kraus operator sum element:

\[
(I \otimes \frac{1}{\sqrt{N}} \sum^N_{i=1} <i| ) U (I \otimes \frac{1}{\sqrt{N}} \sum^N_{i=1} |i> )
\]
\[
= \frac{1}{N} \sum^N_{i=1} P_i
\]
\[
= M
\]

Repeating the procedure $r$ times brings the Kraus operator to $M^r$, and achieves an approximation of a measurement of $\sum_l |\$_l><\$_l|$.

\subsection{Quantum Lightning}

A recent construction of quantum money is Quantum Lightning, which was proposed by Zhandry in 2017 [17]. Quantum Lightning is a formalization of collision free quantum money [16].

Quantum Lightning makes use of non-collapsing collision resistant hash functions. These hash functions are defined by a random set of degree-2 polynomials over $\mathbb{F}_2$.
Quantum Lightning defines the "Lightning Bolt" state $|\text{\faBolt}>$. The verification procedure \textbf{Ver} for bolts is another polynomial time quantum algorithm that either outputs the serial number of a valid bolt, or $\bot$ for invalid bolts. The serial number of a bolt is a deterministic function of the bolt itself, and verification does not perturb the bolt. Bolts are created by a quantum algorithms called "Storms" and denoted $\text{\faCloud}$.

A bolt is generated by the following procedure.

\begin{itemize}

    \item[1.] Randomly choose $n$ random upper-triangular matrices $A_i \in \{0, 1\}^{m \times m}$, and set $\mathbb{A} = \{A_i\}_i$. $\mathbb{A}$ is the public key. Let the hash function $f_\mathbb{A}: \{0, 1\}^m \rightarrow \{0, 1\}^n$ be $f_\mathbb{A}(x) = (x^T \cdot A_i \cdot x)_i$. If we let operations be taken mod 2, this captures general degree 2 functions over $\mathbb{F}_2$.

    \item[2.] Begin with a state $|\phi_0>$.

    \[
    |\phi_0> =
    \]
    \[
    \frac{1}{2^{kn/2}} \sum_{\Delta_1,...,\Delta_k} |\Delta_1,...,\Delta_k>
    \]

    \item[3.] $\Delta$ is defined such that we can run a computation which maps $\Delta = (\Delta_1,...,\Delta_k)$ to an affine space $S_{\Delta}$ s.t. $\forall x \in S, f_\mathbb{A}(x) = f_\mathbb{A}(x + \Delta_j) \forall j$. Then we construct a uniform superposition of elements in $S_{\Delta}$ to yield:
    
    \[
    |\phi_1> = 
    \]
    \[
    \sum_\Delta \sum_{x \in S_\Delta} \frac{1}{2^{kn/2} \sqrt{|S_\Delta|}} |\Delta, x>
    \]

    \item[3.] Compute $f_\mathbb{A}$ in superposition and measure the resulting serial number $y$.
    
    \[
    |\phi_y> \propto 
    \]
    \[
    \sum_{\Delta, x \in S_\Delta : f_\mathbb{A}(x)=y} \frac{1}{|S_\Delta|} |x, \Delta>
    \]
    
    \item[4.] Compute the maps $(x, \Delta_1,...,\Delta_k)$ to $(x, x-\Delta_1,...,x-\Delta_k)$ in superposition. The final state is a bolt:
    
    \[
    |\text{\faBolt}_y> \propto
    \]
    \[
    \sum_{\Delta, x \in S_\Delta : f_\mathbb{A}(x)=y} \frac{1}{|S_\Delta|}
    \]
    \[
    |x, x-\Delta_1,...,x-\Delta_k>
    \]
    \[
    \dot{=} \sum_{x_0,...,x_k : f_\mathbb{A}(x_i)=y \forall i} |x_0,...,x_k>
    \]
    \[
    =(\sum_{x : f_\mathbb{A}(x)=y} |x>)^{\otimes(k+1)}
    \]
    \[
    =|\text{\faBolt}'_y>^{\otimes(k+1)}
    \]

\end{itemize}

To verify a bolt, each of $k+1$ sets of the $m$ registers is verified individually. Each of these "mini verifications" yields either an element in $\{0, 1\}^n$ or $\bot$. Each mini verification must agree, and have the same output for the bolt to be valid.

We assume the mini verification is given $|\phi> = |\text{\faBolt}'_y>$ that corresponds to some serial number $y$. The first step of mini verification is to check if the input state $|\phi>$ is in the space spanned by $|\text{\faBolt}'_z>$ as $z$ varies.
The second step is to evaluate $f_\mathbb{A}$ in superposition in order to learn which of the orthogonal $|\text{\faBolt}'_z>$ states we have. Then, we can measure the result to obtain $y$. For the correct $|\phi> = |\text{\faBolt}'_y>$ this does not perturb the state. This a useful property since it means that a bolt can theoretically be re-used.

Quantum Lightning ensures that any bolt generated by an honest mint is accepted with probability negligibly close to 1. It also ensures that no adversarial bolt generator can generate two coins with the same serial number which would both pass verification.
Zhandry shows in [17] that Quantum Lightning is secure under some assumptions of the multi-collision resistance of a degree-2 hash function. Zhandry also proved that \textit{any} non-collapsing hash function can be used to construct Quantum Lightning, though there are currently no such known hash functions that are proven to be non-collapsing [17].

\section{A Hybrid Payment System}

In February 2019, Coladangelo proposed a payment system based on Quantum Lightning [18]. Quantum Lightning guarantees that no generation procedure can easily create two coins with the same serial number, and no one can clone existing coins. However, the Quantum Lightning scheme itself does not include a mechanism for regulating the generation of valid coins. This mechanism is introduced as a part of Coladangelo's hybrid blockchain payment system.

This payment system is the first use of smart contracts in a quantum setting. Any party can deposit a coin to a smart contract, setting that contract's serial number to match the coin's. The classical certificate that can be found by measuring a valid bolt can also be submitted to a smart contract. If a certificate submitted by a user corresponds to the serial number stored in the smart contract, this means that the user owns bolt. The contract releases all of its coins to the user.

Coladangelo's payment system also considers one of the challenges with practical quantum computing: state decoherence. The downside of using quantum states as coins are that these coins can't be reliably stored for any significant period of time. The payment system makes use of smart contracts to implement a mechanism for lost coin recovery. A user can send a message to a smart contract with a coin whose serial number is the serial number of a coin they have lost. Other users have a time window in which they can challenge this claim by demonstrating that they in fact own the coin with the submitted serial number. If a claim is not challenged, then the coins submitted to the smart contract are returned to the sender of the message, and the serial number of the contract is updated to that of the lost coin.

\subsection{Classical Blockchain}

\subsubsection{The Global Ledger}

The payment system is primarily a classical blockchain, but uses Quantum Lightning as its coins. The classical serial numbers and certificates of the quantum coins are the interface between the quantum and classical elements of the system. The classical blockchain uses a global ledger. The global ledger maintains three sets and the current time:

\vspace{5mm}
\begin{lstlisting}
parties = {} 
contracts = {}
allTransactions = {}
t = 0
\end{lstlisting}
\begin{center}
FIG. 2. Ledger State
\end{center}
\vspace{5mm}

The messages that the global ledger can handle are the following. These are each slightly modified from those given by Coladangelo for clarity and consistency of notation.

\vspace{5mm}

\noindent{\textbf{Register (id, num\_coins) $\rightarrow$ (pid)} allows a user to set their \textit{id} and retrieve their \textit{pid}, which addresses their data in the system. This constitutes the registration of a user with the system. This message can also include a number of coins, which will be set on the registered party's data structure.}

\vspace{5mm}

\noindent{\textbf{Retrieve Party (pid) $\rightarrow$ (id, num\_coins)} can be used to request a registered party's information.}

\vspace{5mm}

\noindent{\textbf{Pay (pid, pid', num\_coins) $\rightarrow$ (trid)} allows user \textit{pid} to send coins to user \textit{pid'}. If \textit{pid} or \textit{pid'} are not valid, simply return $\bot$. If \textit{pid', pid} $\in$ \textit{parties} and \textit{*pid.coins} $>$ \textit{num\_coins}, then:}

\[
*pid'.coins \leftarrow *pid'.coins + num\_coins
\]
\[
*pid.coins \leftarrow *pid.coins - num\_coins
\]
\[
trid \leftarrow |allTransactions| + 1
\]
\[
allTransactions[trid] \leftarrow
\]
\[
(pid, pid', num\_coins, time)
\]

\vspace{5mm}

\noindent{\textbf{Retrieve Transaction (trid) $\rightarrow$ (allTransactions[trid])} allows users to retrieve transaction details.}

\vspace{5mm}

\noindent{\textbf{Smart Contract $\mathbf{(pids, \{(pid, num\_coins_{pid}): pid \in pids\}, circuit, st_0) \rightarrow(cid)}$} allows a user to create a contract. $\{(pid, num\_coins_{pid}): pid \in pids\}$ are initial deposits for each user \textit{pid}. If \textit{pids} $\subseteq$ \textit{parties}, then a new contract can be created.}

\vspace{5mm}

\noindent{\textbf{Retrieve Smart Contract (cid) $\rightarrow$ (params, coins)} allows a user to retrieve the details of a contract if $cid \in contracts$. Otherwise, returns $\bot$.}

\subsubsection{Smart Contracts}

The global ledger handles contract creation through the \textit{Smart Contract} message. However, the contracts themselves handle the most functional contract-related messages.

The contract creation procedure is the following.

\[
cid \leftarrow |contracts| + 1
\]
\[
*cid.params \leftarrow 
\]
\[
(pids, \{(pid, num\_coins_{pid}): pid \in pids\},
\]
\[
circuit, st_0)
\]
\[
*cid.num\_coins \leftarrow 0 
\]
\[
contracts[cid] \leftarrow *cid
\]

Once created, the contract waits for an \textit{Initialize with Coins} message to come from each user $pid \in *cid.params.pids$. If $*pid.coins \geq num\_coins_{pid} \; \forall \; pid \in *cid.params.pids$, then the following occurs.

\[
*pid.coins \leftarrow *pid.coins - num\_coins_{pid} 
\]
\[
\forall \; pid \in *cid.params.pids
\]
\[
*cid.coins \leftarrow *cid.coins + num\_coins_{pid}
\]
\[
\forall \; pid \in *cid.params.pids
\]
\[
st \leftarrow st_0
\]

The smart contract then enters the "execution phase": a loop which repeats until termination. The contract waits for a \textit{Trigger} message from any user $pid \in parties$. This message will provide variables $(pid, witness, time, st, num\_coins)$. If $circuit(pid, witness, time, st, num\_coins)$
\noindent{$\neq \bot$, then the following occurs.}

\[
*pid.coins \leftarrow *pid.coins - num\_coins
\]
\[
*cid.coins \leftarrow *cid.coins + num\_coins
\]
\[
(st, result) \leftarrow circuit(pid, witness, time,
\]
\[
st, num\_coins)
\]

The \textit{result} will indicate how many coins the smart contract should release to user \textit{pid}.

\[
*pid.coins \leftarrow *pid.coins + num\_coins
\]
\[
*cid.coins \leftarrow *cid.coins - num\_coins
\]

\vspace{5mm}

\noindent{\textbf{Initialize with Coins (pid, cid, num\_coins) $\rightarrow$ ()} allows a user to deposit coins into a contract. This is necessary for a contract to enter its execution phase. If $cid \notin contracts$ or $pid \notin *cid.params.pids$, returns $\bot$. Otherwise, the following occurs.}

\[
*cid.coins \leftarrow num\_coins
\]
\[
*pid.coins \leftarrow num\_coins
\]

\vspace{5mm}

\noindent{\textbf{Trigger (pid, cid, witness, time, st, num\_coins) $\rightarrow$ (result)} allows a user to run the circuit associated with a contract with the given parameters. If $cid \notin contracts$, returns $\bot$. $pid$ may be any element of $parties$.}

\subsection{Quantum Lightning Payments}

In this hybrid blockchain scheme, we have not yet mentioned any use of quantum physics. Indeed, the ledger and contracts in Coladangelo's scheme are completely classical. The only element of the system which makes use of quantum effects is the payments system, which is uses Quantum Lightning as a primitive. Coladangelo's payment system defines five procedures:

\begin{itemize}
    \item generate valid quantum coins
    \item make a payment
    \item file a claim for lost coins
    \item prevent malicious attempts at filing claims
    \item trade valid quantum coins for classical coins
\end{itemize}

\subsubsection{Generating Valid Quantum Coins}

The procedure uses the Quantum Lightning Bolt generation procedure $\text{\faCloud}$ and bolt verification procedure \textbf{Ver} both defined by Zhandry [17], which are included in section 4.5.

\[
|\text{\faBolt}> \leftarrow \text{\faCloud}
\]
\[
serial \leftarrow \textbf{Ver}(|\text{\faBolt}>)
\]

Then to use created coins with the blockchain, the \textit{Smart Contract} message may be sent to the global ledger to create a contract. Once this message has been processed, an \textit{Initialize with Coins} message is also sent to the ledger with the $cid$ matching the contract created by the \textit{Smart Contract} message.

\subsubsection{Making a Payment}

A payment involves two parties, the payer $P$ and payee $P'$. The payment procedure involves the following steps.

\begin{itemize}
    \item $P$ sends $|\text{\faBolt}>, cid, serial$ and $num\_coins$ to $P'$.
    \item $P'$ sends a \textit{Retrieve Contract} message to the ledger, retrieving the contract $cid$.
    \item $P'$ accepts the payment if $cid \in contracts$ and \textbf{Ver}($|\text{\faBolt}>$) = $serial$.
\end{itemize}

\subsubsection{Recovering Lost Coins}

In order to recover lost coins, a user $P$ uses the \textit{Trigger} message to cause a smart contract to execute a circuit \textit{BanknoteLost}. This circuit records a request at the current time to $*cid.state$, indicating that a \textit{BanknoteLost} message began to be processed at this time. With the \textit{Trigger} message, user $P$ also provides the serial number $serial$ of the lost coin, and deposits a number of coins $num\_coins$ into the contract $cid$.

During the time $t_{tr}$ that follows, another user $P'$ has the chance to challenge the claim made by $P$ by demonstrating true ownership of the coin with the serial number $serial$. Recall that a bolt of Quantum Lightning is generated using degree-2 polynomial hash function $H$. If $P'$ has access to the bolt, they can verify it through a verification procedure $A$ which will identify only their one, unique bolt  and yield some $m \in \{0, 1\}^\lambda$ such that $H(m) = serial$.

To challenge the claim made by $P$, $P'$ can perform the following new bolt generation and verification:

\[
m \leftarrow A(|\text{\faBolt}>)
\]
\[
|\text{\faBolt}> \leftarrow \text{\faCloud}
\]
\[
serial' \leftarrow \textbf{Ver}(|\text{\faBolt}>)
\]

Then, $P'$ sends a \textit{Trigger} message to the contract with $m$ and $serial'$, running a circuit \textit{ChallengeClaim} which causes the lost coin recovery record to be erased from the contract's state and the coins deposited by $P$ to be returned.

If a claim made by $P$ goes unchallenged for time $t_{tr}$, $P$ can perform the bolt generation and verification procedure and send a \textit{Trigger} message to the contract with the new coin's serial number $serial'$, which will run a circuit \textit{ClaimUnchallenged}. This circuit simply updates the contract's serial number to be the new bolt's and removes the record of the recovery request.

\subsubsection{Trading a Valid Quantum Coin for a Classical Coin}

If a user $P$ owns a quantum coin and wishes to redeem it for classical coins, they can demonstrate ownership by performing $m \leftarrow A(|\text{\faBolt}>)$, and then sending a \textit{Trigger} message which contains $m$ and runs a circuit \textit{RecoverCoins}. The \textit{RecoverCoins} circuit releases the coins which were originally deposited in the contract by $P$ back to $P$.

\section{A Quantum Blockchain Voting Protocol}

Sun, Xin, et al. presented a protocol for voting on a quantum blockchain in January 2019 [19]. Voting can be a suitable application of blockchain technology since the blockchain makes it difficult for participants to falsify claims. Sun, Xin, et al. make use of quantum commitments to design a self-tallying voting protocol.

\subsection{Voting Using Binding Commitments}

The protocol is very simple. Like the other commitment schemes discussed in section 4.2 of this paper, the voting protocol involves two phases. These phases are called the "ballot commitment" and "ballot tallying" phases.

The steps to the ballot commitment phase are the following.

\begin{itemize}
    \item[1.] For each $i \in \{1,...,n\}$ voter $V_i$ generates the $i^{th}$ row of an $n \times n$ matrix of integers $r_{i,1},...,r_{i,n}$ such that $\sum_j r_{io,j} = 0 (mod \; n + 1)$.
    
    \item[2.] For each $i,j$ voter $V_i$ sends $r_{i,j}$ to voter $V_j$ via a quantum secure communication.
    
    \item[3.] Then each voter $V_i$ knows the $i^{th}$ column $r_{1,i},...,r_{n,i}$. $V_i$ computes his/her masked ballot $\hat{v_i} = v_i + \sum_j r_{j,i} (mod \; n + 1)$. $V_i$ commits $\hat{v_i}$ to every tallier of the blockchain via a quantum commitment protocol.
    
\end{itemize}

Ballots are tallied by the following decommitment procedure. $v_i = 0$ is considered a disagreement with the proposal being voted on, $v_i = 1$ is considered an agreement.

\begin{itemize}
    \item[1.] Each voter $V_i$ reveals $\hat{v_i}$ to every tallier of the blockchain by opening his/her commitment.
    \item[2.] The talliers each run the Quantum Honest Success Byzantine Agreement Protocol to reach a consensus on the value of the masked ballot $\hat{v_1},...,\hat{v_n}$.
    \item[3.] The result of the vote is $\sum_i \hat{v_i} = \sum_i v_i (mod \; n + 1)$.
\end{itemize}

\subsection{Handling Dishonest Ballot Talliers}

A Quantum Honest Success Byzantine Agreement Protocol (QHBA)  is used in their voting scheme to identify dishonest ballot talliers.

\vspace{5mm}

\textbf{Definition 5 (Honest success Byzantine agreement protocol (HBA))}
\textit{An honest success Byzantine agreement protocol involves $n$ agents. One of the agents is the sender $S$, and holds an input value $x_s \in D$, where $D$ is a finite domain. A protocol achieves honest success Byzantine agreement if the protocol guarantees the following:
\begin{itemize}
    \item[1.] If the sender is honest, then all honest agents agree on the same output value $y = x_s$.
    \item[2.] If the sender is dishonest, then either all honest receivers abort the protocol, or all honest receivers decide on the same output value $y \in D$. 
\end{itemize}
The protocol is $p$-resilient if the protocol works when less than a fraction of $p$ receivers are dishonest.}

\vspace{5mm}

The QHBA is $\frac{m-2}{m}$-resilient. $m$ is the number of receivers, and is more efficient than a classical HBA protocol when there are many dishonest receivers [19].

\subsubsection{Distribution of Correlated Lists}

The first phase of the QHBA protocol is for correlated lists to be distributed among the agents using quantum secure direct communication.

Let the sender be $S = P_1$. Each agent $P_i \in \{P_{\frac{n}{2}+1},...,P_{n}\}$ is tasked with distributing a list of numbers $L_k^i$ to agent $P_k \in \{P_1,...,P_{\frac{n}{2}}\}$ such that:

\begin{itemize}
    \item[1.] $|L_k^i| = l \; \forall \; k \in \{1,...,n/2\}$, where $l$ is a multiple of 6.
    \item[2.] $L_1^i \in \{0,1,2\}^l$. $\frac{l}{3}$ numbers on $L_1^i$ are 0. $\frac{l}{3}$ are 1. $\frac{l}{3}$ are 2.
    \item[3.] $L_k^i \in \{0,1\}^l \; \forall \; k \in \{2,...,n/2\}$
    \item[4.] $\forall \; j \in \{1,...,l\}$, if $L_1^i[j] = 0$, then $L_2^i[j] = ... = L_{n/2}^i[j] = 0$
    \item[5.] $\forall \; j \in \{1,...,l\}$, if $L_1^i[j] = 1$, then $L_2^i[j] = ... = L_{n/2}^i[j] = 1$
    \item[6.] $\forall \; j \in \{1,...,l\}$, if $L_1^i[j] = 2$, then $ \forall \; k \in \{2,...,m\}$ the probability that $L_k^i[j] = 0$ and that $L_k^i[j] = 1$ are equal.
\end{itemize}

If the number of receivers that report non-compliant lists from a distributor passes a threshold, then that distributor is classified as dishonest.

\subsubsection{Sequential Composition List Formation}

Let the number of honest distributors be $h$. Then the agents perform the following sequential composition.

\[
L_1 = L_1^{n/2 + 1},..., L_1^{n/2 + h}
\]
\[
L_2 = L_2^{n/2 + 1},..., L_2^{n/2 + h}
\]
\[
\vdots
\]
\[
L_{n/2} = L_{n/2}^{n/2 + 1},..., L_{n/2}^{n/2 + h}
\]

\vspace{5mm}

The constructed sequential composition of correlated lists is then $\mathbb{L}$.
\[
\mathbb{L} = (L_1,...,L_{n/2})
\]

\subsubsection{Consensus}

Assuming $h > \frac{n}{2}$, the following procedure can be used to reach a consensus.

First, the sender $S$ sends a binary number $b_{1,k}$ and a list of numbers $ID_{1,k}$ to each receiver $P_k$. $ID_{1,k}$ should indicate all the positions on $L_1$ where $b_{1,k}$ appears to $P_k$. An honest sender will send the same list to all receivers.

Each $P_k$ will compare the $b_{1,k}$ and $ID_{1,k}$ to their list $L_k$. If any honest $P_k$ finds information that is not consistent, then $P_k$ sends $\bot$ to the other receivers. Otherwise, $P_k$ sends $b_{1,k}$ and $ID_{1,k}$ to the other receivers.

After all these messages have been received, each honest $P_k$ checks the following:

\begin{itemize}
    \item[1.] If there were more than two agents who sent binary numbers and lists that were consistent with $L_k$ but some had different binary numbers, $P_k$ outputs $\bot$.
    
    \item[2.] If more than two agents sent the same binary numbers and lists which were consistent with $L_k$, these agents are considered to be honest. $P_k$ outputs the binary number provided by these honest agents.
    
    \item[3.] If more than two agents sent the same binary numbers and lists which were consistent with $L_k$, any other agents are considered dishonest. If all of the dishonest agents sent $\bot$ to $P_k$, then $P_k$ sets $v_k$ to the binary value provided by the honest agents.
    
    \item[4.] In all other cases, $P_k$ outputs $\bot$.
\end{itemize}

Consensus is achieved if at least $\frac{n}{4}$ agents output the same bit value.

Suppose $P_j$ were a dishonest receiver, and $j \geq 2$. $P_j$ would want to send a binary number $b_{j,k}$ and list of numbers $ID_{j,k}$ which was consistent with $L_k$.On $L_j$, there are $\frac{l}{2}$ appearances of $b_{j,k}$. On $L_1$ there are only $\frac{l}{3}$ appearances of $b_{j,k}$. So, there are $\frac{l}{6}$ positions of discord $x$, where $L_1[x] = 2$. If $P_j$ selects a discord position $x$ then with probability $\frac{1}{2}$, $L_k[x] \neq b_{j,k}$. $P_j$ has to avoid all discord positions in order to avoid being identified as dishonest. This has a $(\frac{2}{3})^{\frac{l}{3}}$ probability of success which is very small when $l$ is large. This is rationale behind the checks made by $P_k$ listed above.

\section{Quantum Blockchain Using Entanglement in Time}

Rajan, Del, and Matt Visser published a quantum system design that uses time entanglement to replace the data structure component of blockchain technology [20]. Their approach uses the nonseparability of entangled photons to simulate the links between blocks of data. The approach addresses the issue of blockchain scalability using quantum effects.

Multipartite states like the GHZ entangled state are used to create a chained data structure. In the most trivial example of the approach, the contents of a block might be represented by a pair of bits $r_1 r_2$. These contents are encoded into a temporal Bell state i.e.

\[
|\beta_{r_1r_2}>^{0, \tau} = \frac{1}{\sqrt{2}} (|0^0>|r_2^{\tau}> + (-1)^{r_1}|1^0>|\bar{r_2}^{\tau}>)
\]

As records are created, they are encoded as blocks into temporal Bell states. These photons are created and absorbed at their respective times.

\[
|\beta_{00}>^{0, \tau},
|\beta_{10}>^{\tau, 2\tau},
|\beta_{11}>^{2\tau, 3\tau},
etc.
\]

The bit strings of the Bell states are then effectively "chained" together using entanglement in time. This is accomplished using a fusion process: Bell states are recursively projected into a growing temporal GHZ state. This can be accomplished using an entangled photon-pair production source, a delay line and a Polarizing Beam Splitter (PBS). For example, two Bell states could be fused into the four photon GHZ state:

\[
|\psi+>_{a,b}^{0,0} \otimes |\psi+>_{a,b}^{\tau,\tau}
\]
\[
\xrightarrow[]{\text{delay}} |\psi+>_{a,b}^{0,\tau} \otimes |\psi+>_{a,b}^{\tau,2\tau} = \frac{1}{2} (|h_a^0 v_b^{\tau}> + |v_a^0 h_b^{\tau}>) \otimes (|h_a^{\tau} v_b^{2\tau}> + |v_a^{\tau} h_b^{2\tau}>)
\]
\[
\xrightarrow[]{\text{PBS}} \frac{1}{2} (|h_a^0v_b^{\tau}v_a^{\tau}h_b^{2\tau}> + |v_a^0h_b^{\tau}h_a^{\tau}v_b^{2\tau}>) = |GHZ>^{0,\tau,\tau,2\tau}
\]

The four photons propagate in their own spatial modes and exist at different times, but are time entangled. The state of the blockchain at a given time $t = n \tau$ is:

\[
|GHZ_{r_1r_2...r_{2n}}>^{0,\tau,\tau,2\tau,2\tau,...,(n-1)\tau,(n-1)\tau,n\tau)}
\]
\[
= \frac{1}{\sqrt{2}} (|0^0r_2^{\tau}r_3^{\tau}...r_{2n}^{n\tau}> + (-1)^{r_1}|1^0\bar{r_2}^{\tau}\bar{r_3}^{\tau}...\bar{r_{2n}}^{n\tau}>)
\]

This state contains the classical information $r_1r_2...r_{2n}$. This information can be decoded without measuring the full photon statistics or detecting the photons [21]. The scalability issue is addressed since "any number of photons can be generated with the same setup, solving the scalability problem caused by the previous need for extra resources. Consequently, entangled photon states of larger numbers than before are practically realizable" [22].

\section{Discussion}

Ever since Wiesner's proposal of public-key quantum money in 1960 [9], quantum cryptography has been an active area of research. However, the topic of quantum blockchain is still relatively new. This is clear from a simple search for papers using the keywords "Quantum" and "Blockchain". There has been a steep, almost exponential, increase in publications over the last three years. There are still many open questions in the area. Ongoing research has identified and introduced new unanswered questions. 

We are now beginning to see blockchain technology beong adopted and trusted for critical government processes. For example, a prominent blockchain company ConsenSys Systems [23] has partnered with an initiative created by His Highness Sheikh Mohammad bin Rashid Al Maktoum, Vice President and Prime Minister of the UAE and Ruler of Dubai to use blockchain widely in Dubai [24]. They released a whitepaper at the World Government Summit of 2017 entitled "Building the Hyperconnected Future on Blockchains" [25]. Some companies that believe in the fundamental potential of decentralized governance like ConsenSys have endeavored to bring blockchain technology to areas of society that could be improved in some way by decentralization, with some success. ConsenSys has supported projects in decentralized journalism [26], law [27], digital asset economy [28], supply chains [29] and more.

The longevity of technology that will impact our most important societal structures is worth questioning. There are critical issues with the scaling properties and efficiency of these blockchain technologies which require solutions if any significant distributed ledgers are going to be sustainably implemented. The scaling properties of the immutable distributed data structures used in blockchain networks have been shown to cause demands on memory that are hard to justify. Blockchains that are based on proof-of-work consensus schemes like Bitcoin also encourage massively wasteful resource consumption. Competition in Bitcoin's computationally-intensive scheme coupled with the limitations of the blockchain data structure implementation by Bitcoin also causes issues with throughput of the system as a practical trading platform. The number of transactions that can be processed by Bitcoin is less than seven per second. This is far from the reported 47,000 per second achieved by VISA [30].

These issues have motivated some pushback against the spread of blockchain. China is seeking to stop Bitcoin mining in the country, for example [31]. From a business perspective, blockchain technology is not expected to be viable for full adoption and practical use by mainstream banks for around another ten years [32]. Even so, banks are beginning to implement prototypes and blockchain applications of limited scale now. An IBM survey of 200 global banks [33] showed that 65\% of these banks intended to roll out blockchain-based products between 2016 and 2019.

The majority of blockchain applications that are being developed do not have solutions to the scalability and efficiency issues of their underlying cybersecurity schemes. They are also not prepared to face the challenges of attackers equipped with the quantum computers we expect to see developed within the next ten to twenty years. Companies are laying the groundwork now for technology that will become fundamentally tied to our most important societal structures, and this technology must be poised for viability in the quantum age. This is what has motivated efforts by companies like NXM Labs to introduce autonomous security protocols which can adapt and be securely updated to accommodate new challenges in the future [34]. This is also what has motivated the Quantum Resistant Ledger project [35].

In this review, we have focused on work that attempts to harness quantum computing to improve blockchain technology. These efforts are currently theoretical frameworks, but future quantum computing infrastructure may enable their realization. Their attempts to address the security and efficiency of blockchain crytocurrencies [9, 10, 16, 17], security primitives [11, 12, 13, 14], smart contracts [18], consensus algorithms [19], and data structures [20] will inform and direct the future implementations of quantum blockchain technologies.

Quantum money was arguably one of the first ideas that kickstarted the entire field of quantum cryptography in 1960 [9]. The inherent security of information stored in quantum states using conjugate coding [9] brings  clear benefit to cryptocurrencies. Wiesner's basic scheme inspired the notable work of Bennett and Brassard's BB84 quantum key distribution protocol [10], among many other foundational works in the field. However, the conjugate coding scheme is not perfect for every use. Hence, improvements have been made such as the idea of collision-resistant quantum money introduced by Lutomirski, Andrew, et al [16] and Zhandry's Quantum Lightning framework [17]. Despite its long history, quantum money still comes with some unanswered questions. Zhandry proved that any non-collapsing hash function can be used to construct Quantum Lightning. However, there are currently no known hash functions that are proven to be non-collapsing [17]. It is an open question whether suitable hash functions could be constructed from better-known assumptions, such as the hardness of lattice problems.

Secure communication primitives have been relevant for work in quantum blockchain technology research. We have summarized the key points of foundational work on binding quantum commitment schemes [11, 12, 13, 14]. Binding commitments underlie collision-resistant quantum money, Quantum Lightning. In turn, these are the primitives used by Coledangelo's hybrid quantum blockchain design [18]. Coledangelo's is one of the first hybrid quantum/classical blockchain designs, and notably addresses the issue of decohering quantum money by introducing a novel method for a blockchain participant to prove that they once owned a quantum coin. The design also includes a concept of arbitrary smart contracts much like Ethereum's. Open questions that remain for hybrid blockchain designers include the problem of ensuring the trustworthiness of arbitrary smart contract code and the hardness of the classical blockchain security elements against quantum attacks, among others.

The consensus algorithm presented by Sun, Xin, et al. [19] is a simple approach to consensus which is adapted for use in a quantum blockchain. Their work demonstrates that consensus can be elegantly simple. Comparing the scaling characteristics of their scheme to those of other consensus algorithms may be useful for future works. Rajan, Del, and Matt Visser's blockchain data structure using entanglement in time [20] is an interesting, new perspective on quantum blockchain. Using a partially quantum mechanical data structure for blockchain may enable hybrid blockchain technologies to take advantage of effects such as entanglement swapping using photons  [22], and many violations of local realism. Whether it will become practical from an engineering or economic point of view to harness these effects on a large scale is yet to be determined.

The goal of this review was to provide a summary of current quantum blockchain research that can help to guide future work. There is huge potential for combining quantum resources with  blockchain technology for applications in a variety of sectors including finance, healthcare, manufacturing and other areas where data security in a distributed network is of importance.
We hope that this review will provide a resource to researchers from these different fields and enable further research and development.

\pagebreak

\section{References}

\begin{itemize}

    \item[1.] Fehr, Serge. “Quantum Cryptography.” Foundations of Physics, vol. 40, no. 5, 2010, pp. 494–531., doi:10.1007/s10701-010-9408-4.

    \item[2.] Nofer, Michael, et al. “Blockchain.” Business \& Information Systems Engineering, vol. 59, no. 3, 2017, pp. 183–187., doi:10.1007/s12599-017-0467-3.
    
    \item[3.] Nakamoto, Satoshi. “Bitcoin: A Peer-to-Peer Electronic Cash System.” Bitcoin, Bitcoin, https://bitcoin.org/bitcoin.pdf.
    
    \item[4.] V. Buterin. “Ethereum white paper: a next generation smart contract \& decentralized application platform,” 2013, http://www.theblockchain.com/docs/-Ethereum\_white\_papera\_next\_generation\_smart\_contract\_and\_decentralized\_ap-plication\_platform-vitalik-buterin.pdf
    
    \item[5.] Atzori, Marcella. “Blockchain Technology and Decentralized Governance: Is the State Still Necessary?” SSRN Electronic Journal, 2015, doi:10.2139/-ssrn.2709713.
    
    \item[6.] Yaga, Dylan J., et al. “Blockchain Technology Overview.” NIST, 10 Nov. 2018, https://www.nist.gov/publications/blockchain-technology-overview.
    
    \item[7.] Frankenfield, Jake. “Proof of Stake (PoS).” Investopedia, Investopedia, 11 Aug. 2019, https://www.investopedia.com/terms/p/proof-stake-pos.asp.
    
    \item[8.] Editor. “Blockchain 2.0 - What Is Ethereum [Part 9].” OSTechNix, 9 May 2019, https://www.ostechnix.com/blockchain-2-0-what-is-ethereum/.
    
    \item[9.] Wiesner, Stephen. “Conjugate Coding.” ACM SIGACT News, vol. 15, no. 1, 1983, pp. 78–88.
    
    \item[10.] Charles H. Bennett and Gilles Brassard. "Quantum public-key distribution reinvented." ACM SIGACT News, vol. 18, no. 4, 1983, pp. 51–53.
    
    \item[11.] Unruh, Dominique. “Computationally Binding Quantum Commitments.” Advances in Cryptology – EUROCRYPT 2016 Lecture Notes in Computer Science, 2016, pp. 497–527., doi:10.1007/978-3-662-49896-5\_18.
    
    \item[12.] Ambainis, Andris, et al. “Quantum Attacks on Classical Proof Systems: The Hardness of Quantum Rewinding.” 2014 IEEE 55th Annual Symposium on Foundations of Computer Science, 2014, doi:10.1109/focs.2014.57.
    
    \item[13.] Fehr, Serge. “Classical Proofs for the Quantum Collapsing Property of Classical Hash Functions.” Theory of Cryptography Lecture Notes in Computer Science, 2018, pp. 315–338., doi:10.1007/978-3-030-03810-6\_12.
    
    \item[14.] Unruh, Dominique. “Collapse-Binding Quantum Commitments Without Random Oracles.” Advances in Cryptology – ASIACRYPT 2016 Lecture Notes in Computer Science, 2016, pp. 166–195., doi:10.1007/978-3-662-53890-6\_6.
    
    \item[15.] Bertoni, Guido, et al. “On the Indifferentiability of the Sponge Construction.” Advances in Cryptology – EUROCRYPT 2008 Lecture Notes in Computer Science, 2008, pp. 181–197., doi:10.1007/978-3-540-78967-3\_11.
    
    \item[16.] Lutomirski, Andrew, et al. “Breaking and Making Quantum Money: toward a New Quantum Cryptographic Protocol.” 2009, pp. Innovations in Computer Science - ICS 2010, Tsinghua University, Beijing, China, January 5–7, 2010. Proceedings, 20–31, 978–7-302–21752-7 Tsinghua University Press.
    
    \item[17.] Zhandry, Mark. “Quantum Lightning Never Strikes the Same State Twice.” Advances in Cryptology – EUROCRYPT 2019 Lecture Notes in Computer Science, 2019, pp. 408–438., doi:10.1007/978-3-030-17659-4\_14.
    
    \item[18.] Coladangelo, Andrea. “Smart Contracts Meet Quantum Cryptography.” 2019.
    
    \item[19.] Sun, Xin, et al. “A Simple Voting Protocol on Quantum Blockchain.” International Journal of Theoretical Physics, vol. 58, no. 1, 2019, pp. 275–281.
    
    \item[20.] Rajan, Del, and Matt Visser. “Quantum Blockchain Using Entanglement in Time.” Vol. 1, no. 1, 2018, pp. 3–11.
    
    \item[21.] Megidish, et al. “Quantum Tomography of Inductively-Created Large Multiphoton States.” 2017.
    
    \item[22.] Megidish, et al. “Entanglement Swapping between Photons That Have Never Coexisted.” Physical Review Letters, vol. 110, no. 21, 2013, p. 210403.
    
    \item[23.] ConsenSys. “Get to Know the ConsenSys Mesh.” ConsenSys Media, ConsenSys Media, 26 Feb. 2018, media.consensys.net/get-to-know-the-47-projects-that-make-up-the-consensys-mesh-478b7d3028c1.
    
    \item[24.] “Blockchain for Government: Smart Dubai.” Blockchain for Government: Smart Dubai, consensys.net/enterprise-ethereum/use-cases/government-and-the-public-sector/smart-dubai/.
    
    \item[25.] “ConsenSys Releases Whitepaper At Dubai's World Government Summit.” ETHNews.com, 13 Feb. 2017, www.ethnews.com/consensys-releases-whitepaper-at-dubais-world-government-summit.
    
    \item[26.] “The Civil White Paper.” Civil, civil.co/white-paper/.
    
    \item[27.] “A Free Legal Repository.” OpenLaw, www.openlaw.io/.
    
    \item [28.] Oved, Michael, and Don Mosites. “Swap: A Peer-to-Peer Protocol for Trading Ethereum Tokens.” Whitepaper Database, 21 June 2017, whitepaperd-atabase.com/airswap-ast-whitepaper/.
    
    \item[29.] “Verified Organic and ConsenSys-Backed Treum Launch Ethereum Blockchain Solution to Track and Trace the First Commercial Hemp Crop Planted in Arizona.” IT Supply Chain, 21 June 2019, itsupplychain.com/verified-organic-and-consensys-backed-treum-launch-ethereum-blockchain-solution-to-track-and-trace-the-first-commercial-hemp-crop-planted-in-arizona/.
    
    \item[30.] Vujicic, Dejan, et al. “Blockchain Technology, Bitcoin, and Ethereum: A Brief Overview.” 2018 17th International Symposium INFOTEH-JAHORINA (INFOTEH), 2018, doi:10.1109/infoteh.2018.8345547.
    
    \item[31.] Goh, Brenda, and Alun John. “China Wants to Ban Bitcoin Mining.” Reuters, Thomson Reuters, 9 Apr. 2019, www.reuters.com/article/us-china-cryptocurrency/china-wants-to-ban-Bitcoin-mining-idUSKCN1RL0C4.
    
    \item[32.] Schaus, Paul. “Blockchain Projects Will Pay Off - 10 Years from Now.” American Banker, 2 Dec. 2016, www.americanbanker.com/opinion/blockchain-projects-will-pay-off-10-years-from-now.
    
    \item[33.] Macheel, Tanaya. “Banks Will Start Actually Using Blockchain Next Year: IBM Report.” American Banker, 28 Sept. 2016, www.americanbanker.co-m/news/banks-will-start-actually-using-blockchain-next-year-ibm-report.
    
    \item[34.] “NXM Labs Acclaimed by Frost \& Sullivan for Its Revolutionary Autonomous Security Platform for IoT.” Journal of Engineering, 2019, p. 573.
    
    \item[35.] Matier, Jack, and Pete Waterland. “Quantum Resistant Ledger (QRL).” QRL Whitepaper, Oct. 2016, github.com/theQRL/Whitepaper/blob/master/-QRL\_whitepaper.pdf.
    
\end{itemize}

\end{document}